\documentclass[12pt]{article}
\pdfoutput=1
\usepackage[square,comma,numbers,sort&compress]{natbib}
\usepackage{graphicx,epstopdf,amssymb,amsfonts,amsmath,amsthm,array,
mathrsfs,amscd}
\usepackage[vcentermath]{youngtab}
\usepackage{float}
\usepackage{hyperref}
\usepackage{xcolor}
\newcommand{\pbs}[1]{\let\temp=\\#1\let\\=\temp}
\numberwithin{equation}{section}
\def\be{\begin{equation}}\def\ee{\end{equation}}
\def\cvp{\raise 2pt\hbox{,}} \def\sign{\mathop{\text{sign}}\nolimits}
 \def\tr{\mathop{\text{tr}}\nolimits}

\def\la{\lambda}
 \def\uN{\text{U}(N)}

\def\Tc{T_{\text c}}\def\la{\lambda}\def\mc{m_{\text c}}
\def\Qc{Q_{\text c}}
\def\oD{\text{O}(D)}
\newcommand{\TcValue}{0.0687_2}
\newcommand{\McValue}{0.345_1}
\newcommand{\McZTValue}{0.2125_5}
\newcommand{\Qend}{0.244_6}

\def\plb#1#2#3{{\it Phys.\ Lett.\ }{\bf B #1} (#2) #3}
\def\npb#1#2#3{{\it Nucl.\ Phys.\ }{\bf B #1} (#2) #3}

\def\prl#1#2#3{{\it Phys.\ Rev.\ Lett.\ }{\bf #1} (#2) #3}
\def\jhep#1#2#3{{\it J. High Energy Phys.\ }{\bf #1} (#2) #3}
\def\prd#1#2#3{{\it Phys.\ Rev.\ }{\bf D #1} (#2) #3}
\def\prb#1#2#3{{\it Phys.\ Rev.\ }{\bf B #1} (#2) #3}
\def\prx#1#2#3{{\it Phys.\ Rev.\ }{\bf X #1} (#2) #3}

\def\jmp#1#2#3{{\it J.\ Math.\ Phys.\ }{\bf #1} (#2) #3}
\def\ijmpa#1#2#3{{\it Int.\ J.\ Mod.\ Phys.\ }{\bf A #1} (#2) #3}

\def\imath#1#2#3{{\it Invent math }{\bf #1} (#2) #3}

\def\jpa#1#2#3{{\it J.\ Phys.\ }{\bf A #1} (#2) #3}
\def\ahp#1#2#3{{\it Annales Henri Poincar\'e }{\bf #1} (#2) #3}

\begin{document}
%
%
{\pagestyle{empty}
\parskip 0in
\

\vfill
\begin{center}
{\sffamily\Large\bfseries Phase Diagram of Planar Matrix Quantum Mechanics,}

\bigskip

{\sffamily\Large\bfseries Tensor, and Sachdev-Ye-Kitaev Models}

\vspace{0.4in}


Tatsuo Azeyanagi,$^{1}$ Frank Ferrari,$^{1,2}$ and Fidel I.~Schaposnik Massolo$^{2}$ 
\\

\medskip
${}^1${\it Service de Physique Th\'eorique et Math\'ematique\\
Universit\'e Libre de Bruxelles (ULB) and International Solvay Institutes\\
Campus de la Plaine, CP 231, B-1050 Bruxelles, Belgique}

\smallskip

${}^{2}${\it Center for the Theoretical Physics of the Universe\\
Institute for Basic Sciences (IBS), Seoul, 08826, Republic of Korea}


\smallskip

{\tt tatsuo.azeyanagi@ulb.ac.be, frank.ferrari@ulb.ac.be, fidel.s@gmail.com}
\end{center}
\vfill\noindent

We compute the phase diagram of a $\uN^{2}\times\oD$ invariant fermionic planar matrix quantum mechanics [equivalently tensor or complex Sachdev-Ye-Kitaev (SYK) models] in the new large $D$ limit, dominated by melonic graphs. The Schwinger-Dyson equations can have two solutions describing either a high entropy, SYK black-hole-like phase, or a low entropy one with trivial IR behavior. In the strongly coupled region of the mass-temperature plane, there is a line of first order phase transitions between the high and low entropy phases. This line terminates at a new critical point which we study numerically in detail. The critical exponents are nonmean field and differ on the two sides of the transition. We also study purely bosonic unstable and stable melonic models. The former has a line of Kazakov critical points beyond which the Schwinger-Dyson equations do not have a consistent solution. Moreover, in both models the would-be SYK-like solution of the IR limit of the equations does not exist in the full theory.

\vfill

\medskip
%
\begin{flushleft}
\today
\end{flushleft}
\newpage\pagestyle{plain}
\baselineskip 16pt
\setcounter{footnote}{0}

}

%
\section{\label{IntroSec} Introduction}

A series of recent works \cite{Kitaev,malda1,witten,Ferrari1} combining old ideas developed in the condensed matter literature \cite{condmat}, with new insights in black hole physics \cite{BH1,BH1b,BH2,BH3} and tensor model tech\-no\-lo\-gy \cite{tensor1,tensor2,tensor3,tensor4}, is making it possible, for the first time, to reliably study planar quantum mechanical matrix models in the fully nonperturbative regime. For a partial list of interesting recent papers in this area, see, e.g.,\ \cite{longlist}. The aim of the present work is to initiate a systematic study of the phase diagrams of such models. 

We are going to  focus on the Hamiltonian
\be\label{Hferm} H=ND\tr\Bigl( m\,\psi^{\dagger}_{\mu}\psi_{\mu}+\frac{1}{2}\la\sqrt{D}\,\psi_{\mu}\psi_{\nu}^{\dagger}\psi_{\mu}\psi_{\nu}^{\dagger}\Bigr)\, ,\ee
where $\psi_{\mu}$ and $\psi^{\dagger}_{\mu}$ are $\oD$ vectors of $N\times N$ matrices made of fermionic annihilation and creation operators $(\psi_{\mu})^{a}_{\ b}=\psi^{a}_{\mu\, b}$ and $(\psi^{\dagger}_{\mu})^{a}_{\ b}=(\psi^{b}_{\mu\, a})^{\dagger}$ satisfying canonical anticommutation relations
\be\label{fermcanonical} \bigl\{\psi_{\mu\, b}^{a}, (\psi^{\dagger}_{\nu})^{c}_{\ d}\bigr\}=\frac{1}{ND}\delta_{\mu\nu}\delta^{a}_{d}\delta_{b}^{c}\, .\ee
After taking the usual planar $N\rightarrow\infty$ limit, we are considering the new $D\rightarrow\infty$ limit at fixed $\la$ defined in \cite{Ferrari1}. In this limit, the model is dominated by melonic planar graphs and turns out to be equivalent to the model with quenched disorder investigated by Sachdev in \cite{Sachdev}. Our results thus also immediately apply to models of this type.

We shall also briefly discuss a model of Hermitian bosonic matrices with an unstable potential, equivalent at leading order with the Carrozza-Tanasa-Klebanov-Tarponolsky theory \cite{tensor3,KT}, and another model of Hermitian bosonic matrices with a stable potential, which can be seen as the bosonic truncation of a natural supersymmetric extension \cite{Ferrari2}. 

In this Letter, we focus on the discussion of the physics and on the presentation of the main results. Full technical details, together with a discussion of supersymmetric models, which can be easily constructed in the present framework \cite{Ferrari2}, will be given elsewhere \cite{AFS}.

\section{\label{DiscussionSec} Physical discussion}

Let us start by recalling the three basic steps that allow us to solve models such as \eqref{Hferm} in the large $N$ and large $D$ limits. We first start from perturbation theory to compute physical quantities as a power expansion in the coupling constant $\la$ in terms of Feynman graphs. We then consider the large $N$ and large $D$ limits, which in our case, select the planar melonic graphs described in \cite{Ferrari1}. This truncation of perturbation theory produces a convergent series expansion. In step three, we sum the perturbative series exactly via an appropriate Schwinger-Dyson (SD) equation. This gives access to the ``fully nonperturbative'' regime of the model, by analytic continuation. We thus obtain a nonperturbative description of the physics, but it is conceptually crucial to understand that it relies in a fundamental way on perturbation theory.

A remarkable property of fermionic models like \eqref{Hferm} is that there are two natural but distinct ways to define a perturbative expansion and thus \emph{a priori} two distinct paths to access the nonperturbative physics using the strategy described in the previous paragraph.

The first perturbative regime corresponds to high temperature $T=1/\beta\gg |\la|$ at fixed mass $m$. The fermions then have a high effective thermal mass $\sim 2\pi T$ and, even in the extreme case $m=0$, we get a small effective coupling constant $\la/T$. The zeroth order of the resulting perturbative expansion corresponds to a system in the maximally entropic state, the $T\rightarrow\infty$ limit of the usual thermal density matrix. This may seem rather exotic, at least from the point of view of particle theorists; in particular, for $m=0$, the zeroth order Hamiltonian around which we perturb vanishes. Nevertheless the resulting perturbation theory is perfectly well defined, because the fermionic Hilbert space is finite dimensional with a finite entropy $\ln 2$ per degree of freedom. This ``nonstandard'' perturbation theory is the one used in all the discussions of Sachdev-Ye-Kitaev (SYK)-like models in the literature so far. This is not surprising, because it is the only one available in the original SYK model, which is based on Majorana fermions and corresponds to $m=0$ in our model. 

The second perturbative regime corresponds to large mass $|m|\gg |\la|$ at fixed temperature. The effective coupling constant is then $\la/m$ and the resulting expansion is the familiar perturbation theory around noninteracting fermionic harmonic oscillators of frequency $m$. At zero temperature, the system is in the unique Fock vacuum state and obviously has zero entropy. This remains true after perturbation theory is resummed. This is to be contrasted with the $T\rightarrow 0$ limit of the resummed SYK-like perturbative expansion discussed above, which is well known to display a nontrivial IR behavior with a nonvanishing zero temperature entropy.\footnote{The exception is the so-called $q = 2$ SYK model discussed in \cite{malda1}, which corresponds to a quadratic Hamiltonian.}

We shall see that a genuine phase transition between these two qualitatively distinct regimes is actually possible. Moreover, this phase transition may have a very natural description from a gravitational point of view. Indeed, as with any matrix theory, the Hamiltonian \eqref{Hferm} can be given an abstract brane interpretation in which the operators $\psi^\dag_\mu$ create strings stretching between the branes. Since the zero-point mass of a string is proportional to its length, the
mass term in \eqref{Hferm} is mimicking a separation between the branes by a distance proportional to $m$. In this setup, our phase transition, which corresponds to the appearance of a macroscopic entropy below some critical mass or distance, is very reminiscent of the gravitational collapse of the branes creating a black hole as they come closer together.

To proceed further, we need to briefly review how the solution of the model is obtained. When $N\rightarrow\infty$ and $D\rightarrow\infty$, all the thermodynamical quantities can be expressed in terms of the Euclidean two-point function
\be\label{Gdef1} G(t) =\frac{1}{N}\bigl\langle \tr\text{T}\psi_{\mu}(t)\psi^{\dagger}_{\mu}\bigr\rangle_{\beta}\, .\ee
For example, the total normalized charge (or fermion number) $Q$ and free energy $F$ can be obtained from the relations
\begin{align}\label{basicthermo1} Q=\frac{1}{N^{2}D}\frac{\partial F}{\partial m} &=\frac{1}{N}\bigl\langle\tr\psi_{\mu}^{\dagger}\psi_{\mu}\bigr\rangle_{\beta}=1-G(0^{+})\, ,\\\label{basicthermo2} \frac{1}{N^{2}D}\frac{\partial F}{\partial\la}&=
\frac{1}{2\la}\Bigl(\frac{d}{dt} + m\Bigr) G(0^{+})\, .
\end{align}
The two-point function itself is computed by resumming the leading melonic planar diagrams via the standard SD equations
\begin{align}\label{saddle1} \frac{1}{G_{k}} &= m - 2 i\pi k T + \Sigma_{k}\, ,\\
\label{saddle2}\Sigma(t) &= \la^{2}G(t)^{2}G(-t)\, ,
\end{align}
where the Matsubara-Fourier coefficients $G_{k}$ and $\Sigma_{k}$ are defined for half-integers $k$ in terms of $G$ and $\Sigma$ in the usual way. Note that the same SD equations are obtained in the two perturbative regimes discussed above, because the structure of the Feynman diagrams is the same in both cases. The tree-level propagator is given by
\be\label{treelevel} G^{(0)}(t) = \frac{e^{m(\beta-t)}}{e^{\beta m}+1}\qquad \text{for}\quad 0<t<\beta\ee
and by antiperiodicity $G^{(0)}(t+\beta)=-G^{(0)}(t)$ for other values of $t$. The qualitative difference between the two perturbative expansions comes from the fact that the $T\rightarrow 0$ and $m\rightarrow 0$ limits of $G^{(0)}$ do not commute; depending on the order of the limits, one obtains either the zero temperature SYK propagator $\frac{1}{2}\sign (t)$ or the $m\rightarrow 0$ limit of the zero temperature propagator for massive complex fermions $e^{-m t}\theta (t)$. The existence of these two qualitatively distinct limits has important consequences for the solutions of the SD equations, as we now explain.

The physics at $m=0$, or small $m$, has already been extensively discussed in the literature \cite{Kitaev,Sachdev}. In the strong coupling limit $|\la|\gg T$, we find a nontrivial IR behavior with approximate time reparametrization invariance and a nonvanishing zero temperature entropy. As thoroughly discussed in \cite{Sachdev}, the physics in this regime is very similar to what one would expect for an extremal (at $T=0$) or nearly extremal (at small $T/\la$) black hole. We shall call this the ``high entropy'' (HE) phase and denote the associated solution of the SD equations $G_{\text{HE}}$.

The physics at large $|m|$ and fixed $T$, on the other hand, is quite different. As discussed above, in this regime, the model behaves like a set of weakly coupled matrix harmonic oscillators. In particular, it has an exponentially small entropy at low temperature. In spite of this, as soon as $T>0$, we do expect to find all the standard qualitative features associated with black holes, including a continuous spectrum, quasinormal behavior, and chaos. Very plausibly, the Lyapunov exponent discussed in \cite{BH2} will vary as a function of $m$ and $T$, but will remain strictly positive for all $T>0$. This intuition comes from the fact that these general features are likely to be present in any generic matrix quantum mechanics \cite{BH1b}. This will be consistent with the continuous phase structure described below (see Fig.\ \ref{fig1}). For these reasons, we shall call the large $|m|$, fixed $T$ phase the ``low entropy'' (LE) phase, denoting the associated solution of the SD equations $G_{\text{LE}}$.

At $T=0$, it is actually trivial to check that the LE solution coincides with the tree-level propagator, $G_{\text{LE}}(t;T=0) = G^{(0)}(t;T=0)= \theta(t) e^{-m t}$.
This function solves the SD equations \eqref{saddle1} and \eqref{saddle2} exactly, for all values of $m\geq 0$ and $\la$, because $\theta(t)\theta(-t)=0$ ensures that all melon diagrams vanish. Of course, this simplification is valid only at $T=0$; the LE solution gets nontrivial quantum corrections from the melon diagrams as soon as $T>0$. Thus, we clearly see that the SD equations must have two distinct solutions, at least for low enough temperature and low enough $|m|$, a regime where previous studies in the literature have already demonstrated the existence of the HE phase. It is the competition between the LE and HE solutions in the strongly coupled region of parameter space which governs the phase diagram of the theory; see the next section.

In the case of bosonic models, the above discussion is drastically modified because the only perturbative region is $|m|\gg |\la|$. When $T\gg |\la|$ at fixed $m$, the models become classical but the zero modes remain strongly coupled. In other words, bosonic models do not have an SYK-like perturbation theory. Another way to see this is to note that the $m=0$ perturbation theory for bosons suffers from untamable infrared divergences, whereas, for fermions, there is a gap $\pi T$ and no infrared divergence. The only perturbative phase of bosonic models is thus the analogue of the LE phase of our fermionic theory. The existence of a phase with nontrivial IR behavior for bosonic systems is then \emph{a priori} unclear. We shall see in the last section that, at least for two simple and natural examples, there is no such phase, in spite of the fact that the associated SD equations would naively allow it.

\section{\label{ResultsSec} Summary of the main results}
\begin{figure}
\centering
\def\svgwidth{5in}
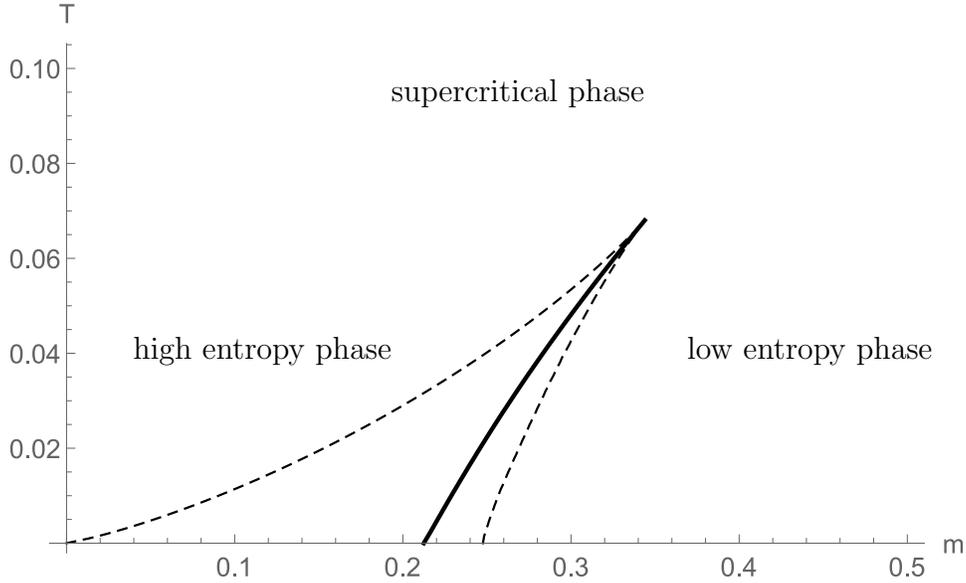
\caption{Phase diagram of the model \eqref{Hferm} in the $(m,T)$ plane, in the units $\la=1$. The plain thick line corresponds to a first order phase transition between high and low entropy phases, whereas the dashed lines delimit the region in which both the high and low entropy solutions exist. There is a nontrivial critical point at $T= \Tc = \TcValue$ and $m = \mc = \McValue$.
\label{fig1}}
\end{figure}
\begin{figure}
\centerline{\includegraphics[width=5in]{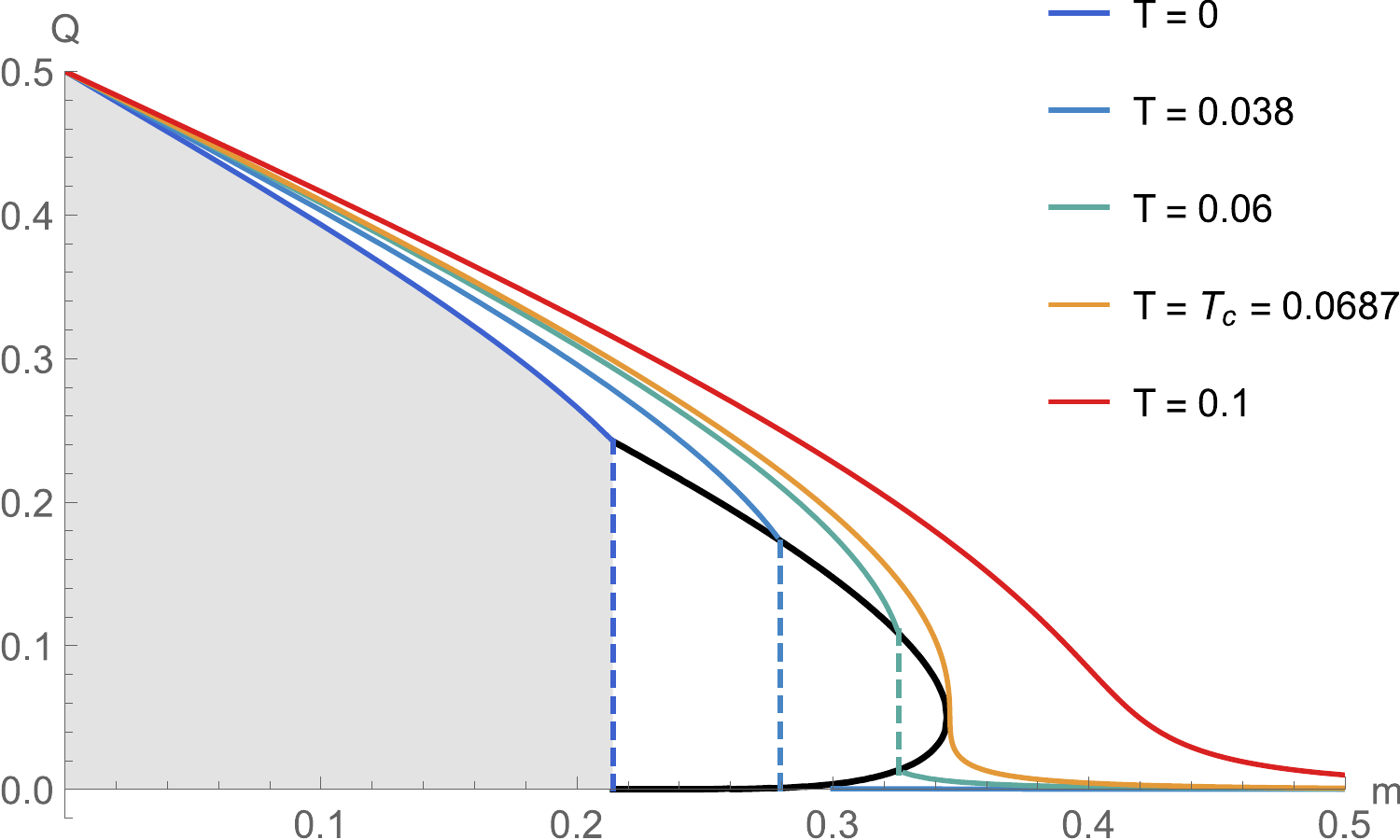}}
\caption{Phase diagram of the model \eqref{Hferm} in the $(m,Q)$ plane, in the units $\la=1$. We have indicated several isothermal curves. For $T<\Tc$, these curves go through the transition region delimited by the thick solid line, where both phases coexist. The area shaded in gray is a forbidden region.
\label{fig2}}
\end{figure}

Let us note that the models for $m$ and $-m$ are equivalent by particle-hole exchange and thus we can choose $m\geq 0$ without loss of generality. Moreover, since only $\la^{2}$ enters at leading order, we can also assume that $\la>0$ and actually that $\la=1$ by choosing the units appropriately. Our main results, obtained by a thorough numerical analysis of Eqs. \eqref{saddle1} and \eqref{saddle2}, whose details will be presented elsewhere \cite{AFS}, are depicted in Figs.\ \ref{fig1} and \ref{fig2}. They  can be summarized as follows.

For $T<\Tc = \TcValue$, the SD equations always have two inequivalent solutions in a certain mass interval, 
$[m_{\text{LE}}(T),m_{\text{HE}}(T)]$, with $ m_{\text{LE}}(\Tc)=m_{\text{HE}}(\Tc)$ and $m_{\text{LE}}(0)=0$. The LE and HE solutions exist for $m\geq m_{\text{LE}}$ and  $m\leq m_{\text{HE}}$, respectively. The boundaries of the existence regions of the two solutions are depicted in dashed lines in Fig.\ \ref{fig1}. The thick black line corresponds to the coexistence line between the two phases, where their free energies are equal. 
When we go from the large mass region to the small mass region along an isotherm at $T<\Tc$, the system thus undergoes a first order phase transition from the LE to the HE phase across the thick line. For example, at $T=0$, this transition occurs at $m = \McZTValue$. For $m > \McZTValue$, the ground state of the system is the Fock vacuum, whereas for $m < \McZTValue$, it is highly degenerate with a nonzero macroscopic entropy of order $N^{2}D$. The coexistence line ends at a critical point $(T = \Tc = \TcValue, m = \mc = \McValue, Q=\Qc = 0.050_{1})$. For $T>\Tc$, there is only one solution to the SD equations, a ``supercritical phase.'' The existence of the critical point allows for a smooth interpolation between the LE and the HE phases, which confirms the intuition that there is no sharp distinction between them. Qualitatively, the difference between the LE and the HE is similar to the difference between the liquid and gaseous phases of water.  

In Fig.\ \ref{fig2}, the phase diagram is depicted in the $(m,Q)$ plane with various isotherms. For $T<\Tc$, the pieces of the isotherms above and below the thick black line correspond to the HE and LE phases, respectively, whereas on the dashed portion these two phases coexist. A notable feature is that there exists a forbidden region in parameter space, shaded in gray. Another interesting property is that charges in the interval $]0,\Qend[$, where $Q = \Qend$ corresponds to the upper left end point of the thick black curve, cannot be realized in a pure phase below a certain temperature.

\begin{figure}[h!]
\centerline{\includegraphics[width=5in]{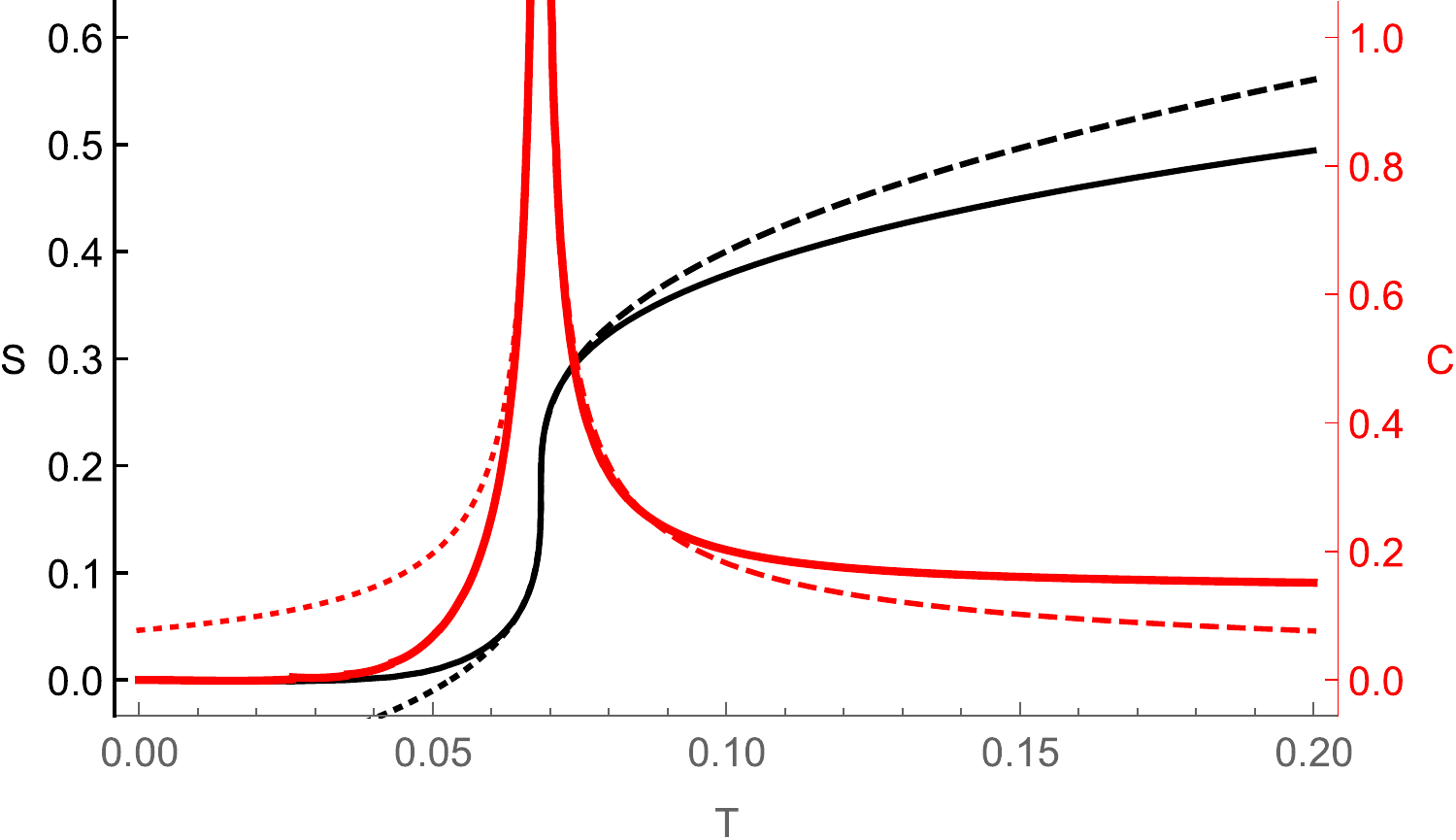}}
\caption{Entropy $S$ (black curve) and heat capacity $C_{m}$ (red curve) at $m=\mc$, as functions of temperature, together with the best power-law fits near $\Tc$ (dashed curves for $T>\Tc$ and dotted curves for $T<\Tc$).
\label{fig3}}
\end{figure}

\begin{figure}[h!]
\centering
\def\svgwidth{5in}
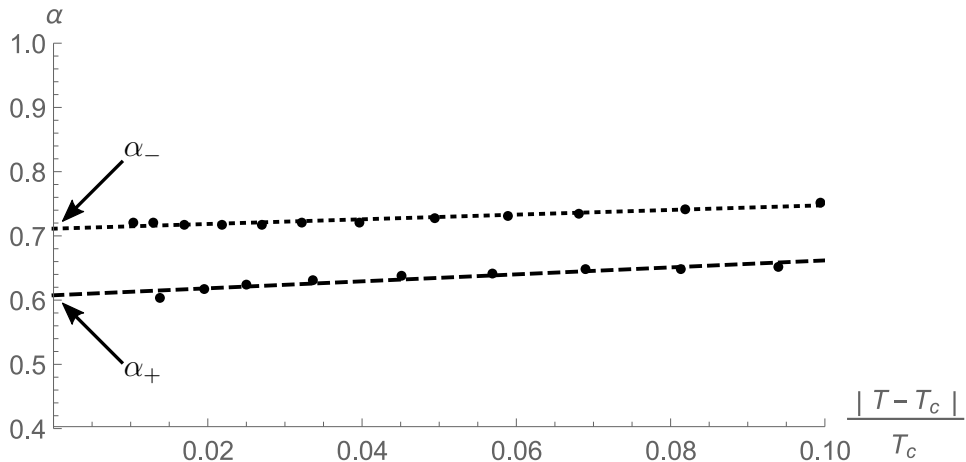
\caption{Plot of \smash{$\alpha=1-\frac{\partial\ln |S(T,\mc)-S(\Tc,\mc)|}{\partial\ln |T-\Tc|}$} as a function of $\smash{\frac{|T - \Tc|}{\Tc}}$, for both $T>\Tc$ (dashed line) and $T<\Tc$ (dotted line). The critical exponents $\alpha_{\pm}$ for the heat capacity are defined by the limits $\alpha_{\pm}=\smash{\lim_{T\rightarrow\Tc^{\pm}}\alpha(T)}$. 
\label{fig4}}
\end{figure}

We have performed an extensive numerical study of the thermodynamical properties around the strongly coupled critical point $(\mc,\Tc)$. We find that various quantities can be very well fitted with a power law, thus defining critical exponents. These exponents are not mean field. Moreover, they are found to be different on both sides of the transition. For example, the heat capacity at fixed $m$, $C_{m}=T\partial S/\partial T$, together with the entropy, is plotted in Fig.\ \ref{fig3}. We clearly get a diverging $C_{m}$ at $\Tc$, with right and left critical exponents given by $\alpha_{+} = 0.61_{1}$ and $\alpha_{-}=0.71_{1}$ (see Fig.\ \ref{fig4}). Other critical exponents are found for the
susceptibility, $\chi = \partial Q/\partial m \propto |T-\Tc|^{-\gamma_{\pm}}$ with $\gamma_{+} = 0.51_1$ and $\gamma_{-}=0.80_{5}$;
for the entropy as a function of mass, $S(\Tc,m)- S(\Tc,\mc) \propto |m-\mc|^{s_{\pm}}$ with $s_{+}=0.31_1$ and $s_{-}=0.37_1$;
for the charge as a function of temperature, $Q(T,\mc)- Q(\Tc,\mc) \propto|T-\Tc|^{q_{\pm}}$ with $q_{+}=0.43_1$ and $q_{-}=0.24_1$;
and for the charge as a function of mass, $Q(\Tc,m)- Q(\Tc,\mc) \propto |m-\mc|^{\tilde q_{\pm}}$ with $\tilde q_{+}=0.26_1$ and $\tilde q_{-}=0.43_1$.
We have also looked at the ``order parameter'' $\Delta Q$, which is the charge difference between the two phases along their coexistence line, with the results $\smash{\Delta Q (T) \propto (\Tc - T)^{\beta}}$ and $\smash{\Delta Q (m) \propto (\mc - m)^{\tilde\beta}}$ for $\beta=0.52_2$ and $\tilde\beta=0.53_1$.


Finding a strongly coupled critical point in the phase diagram of a large $N$ matrix quantum mechanics is an unexpected new feature. Even though our results are compatible with power-law behavior, there is no clear renormalization group (RG) description of the critical point, since our model has no notion of space or locality. In particular, there is no obvious way to write down RG equations in Euclidean signature, so the usual derivations of scaling laws do not seem to apply. This is consistent with our finding of asymmetric critical exponents on the two sides of the transition (see, e.g.,\cite{exponents}).

An important part of the story that we did not investigate is the real time dynamics. On a generic point of parameter space, the correlation functions decay exponentially at large time, consistent with the quasinormal behavior of black holes. Near the critical point, we expect that the quasinormal time scales will diverge and that dynamical critical exponents can be defined. We plan to study this aspect in the future.

\section{\label{BoseSec} Bosonic models}
\begin{figure}
\centerline{\includegraphics[width=5in]{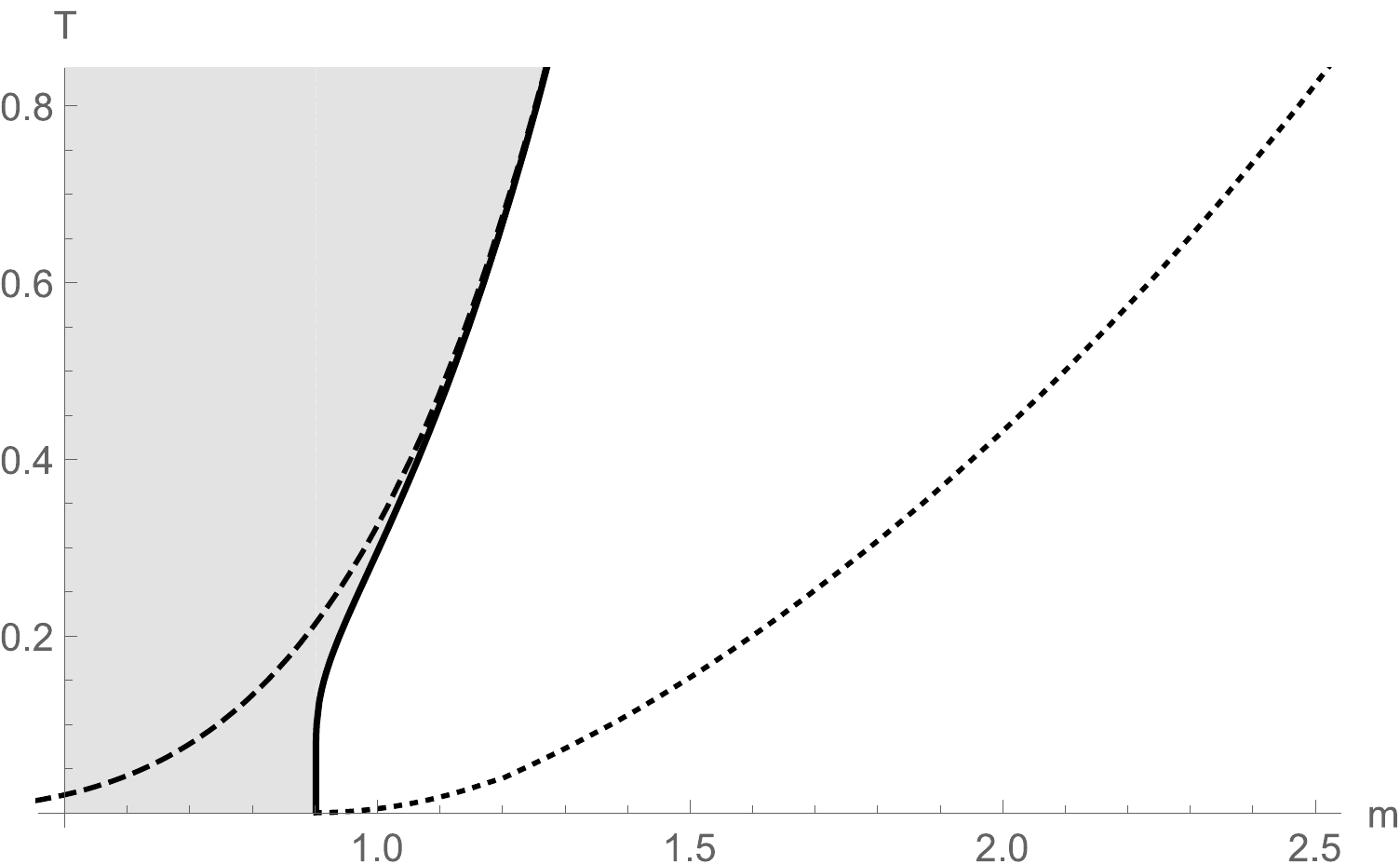}}
\caption{Phase diagram of the model \eqref{pot1} in units $\la=1$. We have plotted the quantum instability line (thick black curve), its classical limit (dashed curve) and the locus of zero entropy for the thermodynamically unfavored solution of the SD equations (dotted curve). This solution has negative entropy to the right of this curve.
\label{fig5}}
\end{figure}

On top of the important qualitative difference between fermionic and bosonic perturbation theories already discussed at the end of the second section, bosonic models may also differ crucially from fermionic ones because of instabilities. Unstable bosonic models are of course not well defined at the full nonperturbative level, but it is well known that they can still be considered in the large $N$ limit, as long as the coupling is not too large. 

We have studied, for example, a model of Hermitian bosonic matrices with unstable potential
\be\label{pot1} ND \tr\Bigl(\frac{m^{2}}{2}X_{\mu}X_{\mu}+\frac{\la^{3}}{4}\sqrt{D}\, X_{\mu}X_{\nu}X_{\mu}X_{\nu}\Bigr)\, .\ee
It is straightforward to write down the SD equations and check that they have the following properties:

\noindent\ \ (i) They have no real solution in a large strongly coupled region of the $(m,T)$ plane, shaded in gray in Fig.\ \ref{fig5} and bounded by an instability curve.\\\hphantom
\ \ (ii) To the right of the instability curve, the SD equations have two real solutions. This is reminiscent of the fermionic case, but the physics is actually very different. One of the solutions is the LE phase, which corresponds to the standard perturbative solution found at large $m$, fixed $T$. It turns out that this solution is always thermodynamically favored, so there is no phase transition. Moreover, the other solution is unphysical, at least for a large region of parameter space where its entropy is found to be negative (to the right of the dotted line in Fig.\ \ref{fig5}).\\\hphantom
\ \ (iii) The two solutions coincide on the instability curve. This phenomenon is a quantum mechanical version \cite{Ferraridoublescale} of the famous Kazakov critical points found in zero-dimensional matrix models \cite{kazakov}. At very high temperature, which is the classical regime, the description is particularly simple: the SD equations reduce to a degree four algebraic equation for the bosonic zero mode and the instability line corresponds to the locus where the two real roots of this equation coincide (dashed line in Fig.~\ref{fig5}).\\\hphantom
\ \ (iv) The model does not have a HE, SYK-like phase. The LE solution has the usual trivial IR behavior, with vanishing zero-temperature entropy; the IR behavior of the unphysical solution can be studied as well and shown to be inconsistent with the SYK properties. The surprising thing about this result is that the SD equations can actually be naively solved in the IR by the usual scale-invariant ansatz for the two-point function at $T=0$, $G(t)=b/|t|^{2\Delta}$ with $\Delta = \frac{1}{4}$ and $b^{-4}=4\pi\la^{6}$. However, this ansatz never corresponds to a well-defined solution of the full SD equations.\footnote{This shows, in particular, that the uncolored bosonic tensor model studied in \cite{KT}, which is equivalent to the case of $m=0$ in our model, cannot have a nontrivial SYK phase.}

Finally, let us note that the instability that we have described in bosonic matrix models could be interpreted differently in the context of quenched disorder models as a sign of spontaneous replica symmetry breaking.

We have also studied a stable purely bosonic model with potential $$ND \tr\Bigl(\frac{m^{2}}{2}X_{\mu}X_{\mu}+\la^{4} D\, X_{\rho}X_{\mu}X_{\rho}X_{\sigma}X_{\mu}X_{\sigma}\Bigr)\, ,$$ which can be solved at leading order by rewriting the potential in a form similar to \eqref{pot1} using an auxiliary field $F_{\mu}\sim X_{\rho}X_{\mu}X_{\rho}$. We find no phase transition or SYK-like solution for this model, again contrary to what a naive scale-invariant ansatz analysis would suggest.

\subsection*{Acknowledgements}

This research is supported in part by the Belgian Fonds National de la Recherche Scientifique FNRS (convention IISN 4.4503.15), the F\'ed\'eration Wallonie-Bruxelles (Advanced ARC project ``Holography, Gauge Theories and Quantum Gravity'') and by IBS-R018-D2. F.~S.~M.\ would like to thank the Service PTM at ULB, Brussels, for its kind hospitality during the completion of this work. We would like to thank S. Sachdev for useful comments on a manuscript of this work. F.~F.\ would like to thank T.~Damour, V.~Pestun and E.~Rabinovici for organizing a wonderful workshop ``Black Holes, Quantum Information, Entanglement and All That'' at the IH\'ES in Bures-sur-Yvette, France (29 May -- 1 June 2017), where this work was first presented.

\end{document}